\documentclass[twocolumn,prl]{revtex4}
\usepackage{graphicx}
\usepackage{amsmath}
\usepackage{amsfonts}

\begin{document}
\title{Boundary-induced nonequilibrium phase transition into an absorbing state}

\author{A. C. Barato and H. Hinrichsen}
\affiliation{Universit\"at W\"urzburg,
	 Fakult\"at f\"ur Physik und Astronomie,
         97074 W\"urzburg, Germany}

\parskip 1mm
\def\d{{\rm d}}
\def\Ps{{P_{\scriptscriptstyle \hspace{-0.3mm} s}}}
\def\MF{{\mbox{\tiny \rm \hspace{-0.3mm} MF}}}

\begin{abstract}
We demonstrate that absorbing phase transitions in one dimension may be induced by the dynamics of a single site. As an example we consider a one-dimensional model of diffusing particles, where a single site at the boundary evolves according to the dynamics of a contact process. As the rate for offspring production at this site is varied, the model exhibits a phase transition from a fluctuating active phase into an absorbing state. The universal properties of the transition are analyzed by numerical simulations and approximation techniques.
\end{abstract}

\maketitle


Nonequilibrium phase transitions differ significantly from ordinary transitions at thermal equilibrium. For instance, under non-equilibrium conditions continuous phase transitions may occur even in one-dimensional systems. A well-known example is the contact process for epidemic spreading~\cite{MarroDickman99}, where diffusing particles multiply at rate $\lambda$ and self-annihilate at rate $1$. Depending on $\lambda$, the contact process is either able to sustain a positive stationary density of particles or it approaches a so-called absorbing state without particles from where it cannot escape. The active and the absorbing phase are separated by a continuous transition belonging to the universality class of directed percolation (DP)~\cite{Hinrichsen00,Odor04,Lubeck04}, which plays a paradigmatic role like the Ising model in equilibrium statistical mechanics. Recently, the critical behavior of DP was confirmed experimentally for the first time by Takeuchi \textit{et al.}~\cite{TakeuchiEtAl07}.

As continuous phase transitions involve long-range correlations, boundary effects may play an important role. In the context of absorbing phase transitions previous studies focused primarily on DP confined to parabolas~\cite{KaiserTurban94,KaiserTurban95}, active walls~\cite{HinrichsenKoduvely98}, as well as absorbing walls and edges~\cite{Froedh98a,Froedh01a}. Although such boundaries influence the dynamics deep in the bulk, the universality class of the bulk transition is not changed inherently, rather it is extended by an additional exponent describing the order parameter near the boundary. A completely different situation is encountered in systems where boundary effects induce a new type transition which would be absent without the boundary~\cite{HenkelSchuetz94}. Such \textit{boundary-induced phase transitions} have been studied for example in models for diffusive transport~\cite{Krug91,Schuetz00} and traffic flow~\cite{PopkovEtAl01}.

In this Letter we present an example of a boundary-induced phase transition from a fluctuating phase into an absorbing state. To this end we consider a simple one-dimensional model, where the leftmost site evolves in the same way as in the contact process while particles in the bulk diffuse according to a symmetric exclusion process. Varying the rate for offspring production at the leftmost site the model exhibits a non-equilibrium phase transition from a fluctuating active phase into an absorbing state with a non-trivial critical behavior. A similar problem with catalytic creation and pair-annihilation at a single site in the center and diffusion in the bulk was studied in \cite{DeloubriereWijland02} by field-theoretic methods.  

\begin{figure}[b]
\includegraphics[width=72mm]{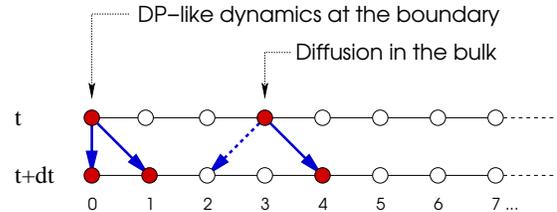}
\vspace{-2mm}
\caption{Definition of the model: Particles diffuse on a semi-infinite one-dimensional chain according to a symmetric exclusion process. The only exception is the leftmost site, where particles multiply and annihilate as in a contact process.}
\label{fig:1d} 
\end{figure}

\paragraph{Definition of the model:}

\begin{figure}[ht]
\includegraphics[width=73mm]{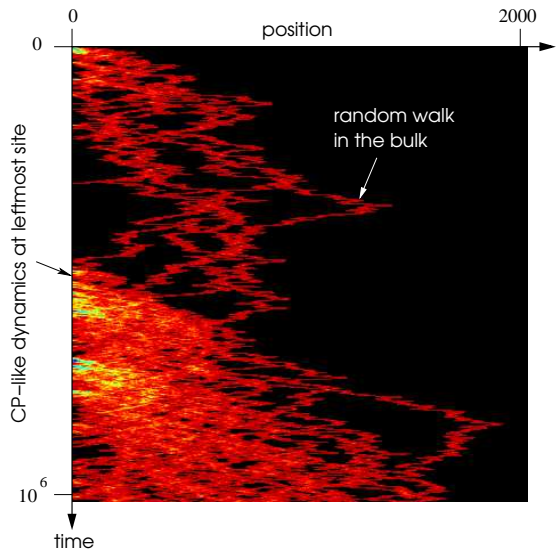}
\vspace{-2mm}
\caption{Typical temporal evolution of the model at criticality, starting with a single particle at the leftmost site. The color scale visualizes the particle density. At the leftmost site one observes intermittent bursts of activity.}
\label{fig:demo} 
\end{figure}

The model is defined on an semi-infinite one-dimensional chain of sites $i=0,1,2,\ldots$ which are either empty ($s_i=0$) or occupied by a particle ($s_i=1$) (see Fig.~\ref{fig:1d} ). Starting at time $t=0$ with a single particle at the origin ($s_i=\delta_{i,0}$) the model evolves by random-sequential updates as follows. For each update one of the particles is randomly selected. If the selected particle is located at the leftmost site $i=0$, it undergoes the same dynamics as in a standard contact process, namely:

\begin{figure*}[t]
\includegraphics[width=175mm]{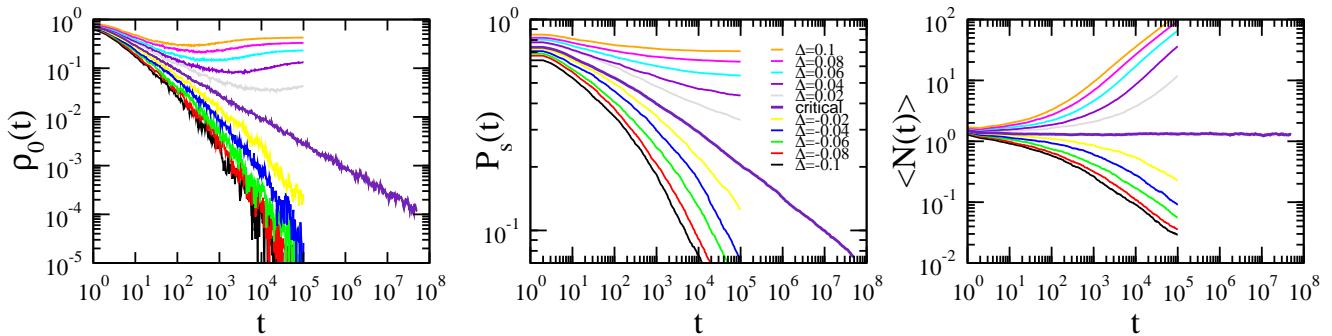}
\vspace{-2mm}
\caption{Numerical simulation of the process starting with a single active site at the origin for different values of $\Delta= p-p_c$. The graphs show the density at the leftmost site $\rho_0(t)$, the survival probability $\Ps(t)$, and the number of particles $\langle N(t) \rangle$ averaged over all runs. }
\label{fig:decay} 
\end{figure*}

\begin{enumerate}
\item[(a)] With probability $p=\lambda/(1+\lambda)$ a new particle is created at the right neighbor, provided that this site is empty. This can be done by setting $s_1:=1$.
\item[(b)] Otherwise, the particle at the leftmost site is destroyed by setting $s_0:=0$.
\end{enumerate}
Else, if the selected particle is \textit{not} located at the origin, it diffuses according to a symmetric exclusion process, i.e., it jumps to a randomly chosen nearest neighbor, provided that the target site is empty. As usual in models with random-sequential dynamics, each attempted update corresponds to a time increment of $1/N(t)$, where $N(t)$ is the actual number of particles. On a computer the dynamical rule defined above can be implemented efficiently by using a dynamically generated list of particle coordinates, eliminating possible finite-size effects.

\paragraph{Phenomenological properties:}

In the bulk the symmetric exclusion process preserves the number of particles, whereas this conservation law is violated at the boundary, where offspring production and removal compete one another. For $p=0$ the leftmost site acts as a sink where particles disappear, thereby depleting the whole system diffusively until the dynamics reaches the absorbing state without particles. On the other hand, for $p=1$ the leftmost site is permanently occupied, providing a steady source of particles at the left boundary so that the system approaches a fully occupied stationary state. In between it turns out that the (infinite) system is able to maintain a non-vanishing stationary density of particles even for finite values of $p$ down to a well-defined critical threshold~$p_c$.

Starting with a single particle at the boundary, the process evolves as follows. Initially the particle at the leftmost site either disappears or it creates another particle at its right neighbor. As soon as this freshly created particle diffuses away into the bulk, the particle at the leftmost site may create and send out further particles until it disappears by spontaneous removal. The average number of newly created particles depends on $p$ and is of order $1$ in the stationary state. 

Each created particle performs a one-dimensional random walk in the bulk, which in one dimension is bound to return to the origin after finite time. The returning particles may either disappear or release another bunch of particles. As demonstrated in Fig.~\ref{fig:demo}, particles are not created continuously but in form of intermittent bursts. Apparently these irregular bursts are responsible for the nontrivial properties of the model.

\paragraph{Seed simulations at criticality:}

The simplest order parameter describing the phase transition is the occupation probability $\rho_0(t)=\langle s_0(t)\rangle$ of the leftmost site averaged over many independent runs. For small $p\ll 1$ this quantity is dominated by the first-return probability of a one-dimensional random walk which is known to decay with time as $t^{-3/2}$ (see e.g.~\cite{Redner01}). This power-law decay characterizes the inactive phase of the system. Contrarily, for large values of $p$, the returning particle is likely to multiply frequently, flooding the bulk of the system with freshly created particles and thereby maintaining a constant non-zero density. In between we find a phase transition located at the critical point (see Fig.~\ref{fig:decay})
\begin{equation}
p_c \;=\; 0.74435(15)\,,
\end{equation}
at which $\rho_0(t)$ decays as 
\begin{equation}
\rho_0(t)\sim t^{-\alpha}\,,\qquad \alpha= 0.50(1)\,,
\end{equation}
suggesting the exact value $\alpha=1/2$.

Another well-known order parameter is the survival probability $\Ps(t)$ to find at least one particle in the entire system at time~$t$. At the transition this quantity is found to decay algebraically as
\begin{equation}
\label{Survival}
\Ps(t) \;=\; t^{-\delta}\,,  \qquad \delta \approx 0.15(2).
\end{equation}
This estimate shows a slight systematic drift and may be compatible with the rational value $\delta=1/6$.

Finally, it is useful to study the number of particles $\langle N(t) \rangle$ averaged over all runs. For small $p$, this quantity decreases as $\langle N(t) \rangle\sim t^{-1/2}$, while for large values of $p$ one finds an algebraic increase $\langle N(t) \rangle\sim t^{+1/2}$ because of the diffusive bulk dynamics. At the transition $\langle N(t) \rangle$ stays almost constant close to $1$, suggesting a vanishing exponent
\begin{equation}
\langle N(t) \rangle \sim t^\theta\,,\qquad \theta=0.
\end{equation}
Note that $\langle N(t) \rangle$ is averaged over \textit{all} runs. If the number of particles was averaged over \textit{surviving} runs it would actually grows as $t^\delta$.

\begin{figure}
\includegraphics[width=87mm]{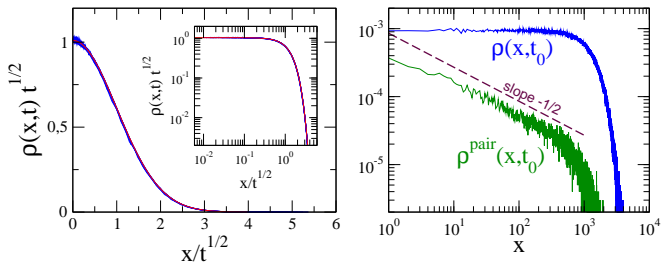}
\vspace{-2mm}
\caption{Left: Data collapse of the rescaled profiles of the particle density at criticality for $t_0=64,128,\ldots,8192$ (blue) compared to a Gaussian distribution (red). Inset: The same data collapse in a double-logarithmic representation. Right: Density of particles (blue) and pairs (green) at $t_0=10^6$, demonstrating the presence of correlations which decay in space as $x^{-1/2}$, indicating that $\beta/\nu_\perp=1/2$.}
\label{fig:profile} 
\end{figure}

We also determined the density profile $\rho(x,t)$ in the bulk (see Fig.~\ref{fig:profile}). At criticality this profile turns out to be an almost perfect Gaussian distribution, obeying the scaling form $\rho(x,t) = t^{-1/2} f(x^2/t)$. This indicates a simple diffusive behavior in the bulk. On the other hand, the density of neighboring \textit{pairs} of particles is not a Gaussian, demonstrating that the random walks are mutually correlated.

\paragraph{Off-critical seed simulations:}

As can be seen in Fig.~\ref{fig:decay}, for $p>p_c$ the average density at the boundary first decreases algebraically, goes through a minimum, then increases again until it reaches a stationary value. Surprisingly, the time at which the minimum is reached scales roughly as $(p-p_c)^{-3}$ while the stationary value is reached at a typical time that scales as $(p-p_c)^{-4}$. Therefore, it is impossible to produce a data collapse by plotting $\rho(t)/(p-p_c)^\beta$ versus $t (p-p_c)^{\nu_\parallel}$. However, collapsing the crossover from increase to saturation, one would consistently get the exponents $\beta=1$ and $\nu_\parallel=4$.

\paragraph{Homogeneous initial state:}

Starting with a fully occupied lattice at criticality, the special dynamics at the leftmost site gradually depletes the system, leading to a slow decay of the particle density at the boundary. In numerical simulations we find that the density decays slowly as $t^{-\delta}$, where $\delta=0.15(2)$ is the same exponent as in Eq.~(\ref{Survival}) which describes the survival of a cluster generated from a single seed. As will be explained in a forthcoming publication, this can be traced back to a duality of the two situations under time reversal.

\paragraph{Mean field analysis:}

In a simple mean field approximation the $n$-site probability distribution is approximated by the product of $n$ single-site probabilities, neglecting correlations. Defining $\eta_i(t)$ as the first moment of the probability distribution at site $i$, the mean field equations read
\begin{eqnarray}
\frac{d\eta_0}{dt}&=& -(1-p)\eta_0+ \frac{1}{2}\eta_1(1-\eta_0),
\label{eqmf1}\\
\frac{d\eta_1}{dt}&=& p\eta_0(1-\eta_1)+ \frac{1}{2}\left(\eta_{2}+\eta_{0}\eta_1-2\eta_{1}\right),
\label{eqmf2}\\
\frac{d\eta_i}{dt}&=& \frac{1}{2}\left(\eta_{i+1}+ \eta_{i-1}-2\eta_{i}\right)  \mbox{ for } i=2,3,\ldots\,.
\end{eqnarray}
Solving these equations for $\eta_i(0)=\delta_{i,0}$ we find the critical point is $p_c^\MF=1/2$, where equation (\ref{eqmf2}) reduces to a diffusion equation, reproducing the critical exponents $\beta^\MF=1$ and $\alpha^\MF=1/2$. However, starting with a fully occupied lattice $\eta_i(0)=1$ one gets a decay $\eta_0(t)\sim t^{-1/4}$, differing from the simulation result~\footnote{In a pair mean field approach, where correlations between the first two sites are taken into account, one obtains a set of four equations instead of Eqs.~(\ref{eqmf1}) and~(\ref{eqmf2}). With this approximation we obtain the same exponents and $p_c^\MF\approx0.634$, which is, as expected, closer to the critical threshold of the full model.}.

\paragraph{Possible relation to a non-Markovian process:}

To understand the transition of the model from a different point of view, let us now adopt the perspective of the leftmost site. If this site is active, it may create new particles, sending them out for random walk in the bulk. From the prospect of the leftmost site the specific trajectory of this random walk does not matter, the only question of interest will be at which time the particle returns to the origin.

Let us now assume that the diffusing particles in the bulk do not interact. For a symmetric exclusion process this approximation is justified if the particle densities are sufficiently small. With this approximation a particle emitted at the leftmost site will return after a time $\Delta t$ which is distributed algebraically as~\cite{Redner01}
\begin{equation}
\label{WaitingTimeDistribution}
P(\Delta t)\sim (\Delta t)^{-3/2}\,.
\end{equation}
Following \cite{DeloubriereWijland02} the problem can be reformulated as a \textit{single-site process} with a \textit{non-Markovian} dynamics. Let $s(t)=0,1$ denote the occupancy of a single site at time $t \in \mathbb{N}$, which can be implemented as a one-dimensional array {\tt s[t]} on a computer. The array is initialized by $s(t):=\delta_{t,0}$, corresponding to a single particle at the boundary. The single-site model then evolves according to the following dynamical rules:
\begin{enumerate}
\item Select the lowest $t$ for which $s(t)=1$.
\item With probability $\mu$ generate a waiting time $\Delta t$ according to the distribution~(\ref{WaitingTimeDistribution}), truncate it to an integer, and set $s(t+\Delta t):=1$.
\item Otherwise (with probability $1-\mu$) set $s(t):=0$.
\end{enumerate}
These steps are repeated until the system enters the absorbing state or $t$ exceeds a predetermined maximal time.

Simulating this non-Markovian single-site process using a dynamically generated list we can go up to $10^{12}$ time steps, finding the critical point $\mu_c=0.57426(1)$, the correct exponent $\alpha=0.500(5)$, as well as a consistent exponent for the survival probability $\delta=0.16(1)$. Moreover, the off-critical properties of the original model are faithfully reproduced. This suggests that the non-Markovian process defined above may be even equivalent to the original model regarding its asymptotic critical behavior. This is surprising since the approximation ignores the exclusion principle of the random walkers in the bulk. 

\paragraph{Relation to a non-Markovian Langevin equation:}

Let us finally describe the single-site process in the continuum limit. As shown in previous studies (see e.g.~\cite{Hinrichsen07} and references therein), a non-Markovian dynamics by algebraically distributed waiting times $P(\Delta t) \sim \Delta t^{-1-\kappa}$ is generated by so-called fractional derivatives $\partial_t^\kappa$ which are defined by
\begin{equation}
\label{eq:IntegralTime}
\partial_t^\kappa \, \rho(t)  \;=\;  \frac{1}{\mathcal{N}_\parallel(\kappa)}
\int_0^{\infty} {\rm d}t' \, {t'}^{-1-\kappa} [\rho(t)-\rho(t-t')]\,,
\end{equation}
where $\kappa\in[0,1]$ and $\mathcal{N}_\parallel(\kappa)=-\Gamma(-\kappa)$ is a normalization constant. This suggests that the non-Markovian single-site model may be effectively described by a DP-like Langevin equation without space dependence, in which the local time derivative is replaced by a fractional derivative with $\kappa=1/2$ generating temporal Levy flights:
\begin{equation}
\label{Langevin}
\partial_t^{1/2} \rho(t) = a \rho(t) - \rho(t)^2 + \xi(t)\,.
\end{equation}
Here the parameter $a$ plays the role of $\mu-\mu_c$, the second term accounts for the fact that the leftmost site cannot be activated twice, and $\xi$ is a multiplicative noise with correlations $\langle \xi(t) \xi(t')\rangle=\rho(t)\delta(t-t')$. Dimensional analysis confirms that the noise is relevant under temporal rescaling, supporting the expectation that the model exhibits a non-mean-field properties. We note that this Langevin equation can be converted by standard techniques into a Fokker Planck equation of the form
\begin{equation*}
\partial_t^{1/2} P(\rho,t) = -\partial_\rho \Bigl[(a\rho -\rho^2) P(\rho,t)\Bigr] 
+ \frac12 \partial_\rho^2\Bigl[ \rho P(\rho,t) \Bigr]\,.
\end{equation*}
This equation can be decoupled by a Laplace transformation in $t$ but so far we were not able to solve the resulting ordinary differential equations analytically.

\paragraph{Towards a scaling picture:}

Based on these results we conjecture that the full model is described by an order parameter field at the boundary with the exponent $\beta=1$ and a response field with the same exponent $\beta'=1$. The diffusive dynamics in the bulk, characterized by a dynamical exponent $z=2$, is slaved to the boundary dynamics and thus it does not induce additional order parameter exponents. Moreover, the numerical results indicate that the scaling exponents are given by $\nu_\parallel=4$ and $\nu_\perp=2$. The decay exponent $\alpha$ describes a two-point function and hence it is given by $\alpha=(\beta+\beta')/\nu_\parallel=1/2$, while the slip exponent is consistently described by the hyperscaling relation $\theta = (\beta+\beta'-\nu_\perp)/\nu_\parallel=0$. The only exponent, which is so far not explained within this picture, is the survival exponent $\delta \approx 0.16$. We believe that this can be traced back to the non-Markovian character of Eq.~(\ref{Langevin}), which requires a new understanding of the survival probability.

To summarize, we have studied a model that exhibits a novel class of boundary-induced phase transition from a fluctuating phase into an absorbing state. This is probably the simplest non-trivial absorbing phase transition, much simpler than ordinary DP, but nevertheless exhibiting properties which cannot be explained within mean field theory.

Financial support by the Deutsche Forschungsgemeinschaft (HI 744/3-1) is gratefully acknowledged. We thank M. A. Mu\~noz for pointing out Ref. \cite{DeloubriereWijland02}.
\vspace{-2mm}



\end{document}